\documentclass{article}
\usepackage{graphicx}

\usepackage[eandd, final]{neurips_2026} 

\usepackage[utf8]{inputenc} 
\usepackage[T1]{fontenc}    
\usepackage{xcolor}
\usepackage[
    colorlinks=true,
    linkcolor=blue!55!black,
    citecolor=green!45!black,
    urlcolor=magenta!55!black,
    filecolor=magenta!55!black,
    pdfborder={0 0 0}
]{hyperref}

\usepackage{url}            
\usepackage{booktabs}       
\usepackage{amsfonts}       
\usepackage{nicefrac}       
\usepackage{microtype}      
\usepackage{xcolor}         
\usepackage{booktabs}
\usepackage{pdflscape}
\usepackage{multirow}
\usepackage{tabularx}
\usepackage[most]{tcolorbox}
\usepackage[subtle]{savetrees}

\renewcommand{\answerTODO}[1][]{\textcolor{red}{\bf [TODO]}}
\renewcommand{\justificationTODO}[1][]{\textcolor{red}{\bf [TODO]}}

\title{PitchBench: Measuring Pitch Hearing in Audio-Language Models }

\author{%
  Milan Liessens Dujardin\textsuperscript{1 *} \quad
  Song-Ze Yu\textsuperscript{1 *} \\[3pt]
  \textbf{Craver Corbyn Thomas-Smith\textsuperscript{2} \quad
  David M. Chan\textsuperscript{1} \quad
  Karina Nguyen\textsuperscript{2}} \\[3pt]
  \textsuperscript{1}University of California, Berkeley
  \quad
  \textsuperscript{2}Thoughtful Lab
  \qquad \\[4pt]
  {\scriptsize \textsuperscript{*}Equal contribution}
}
\begin{document}

\maketitle

\begin{abstract}

Audio-language models (ALMs) are increasingly used in real-world applications that require music understanding, including tutoring, captioning, recommendation, and production. 
Yet existing benchmarks rarely evaluate a fundamental component of this understanding: pitch hearing. Pitch underpins the identification of notes, chords, and melodies and the grounding of captions in musical events, but it remains unclear how reliably ALMs perceive fine-grained pitch across sound sources, acoustic conditions, and response formats.

We introduce \textbf{PitchBench}, an evaluation suite comprising 28 tasks that assess absolute and relative pitch perception under variations in loudness, note duration, sound source, time stretching, background noise, and other acoustic conditions. Tasks range from identifying isolated pitches to tracking voices in four-part musical textures. Evaluating frontier ALMs, we find that pitch hearing remains highly unreliable. Models perform poorly across settings, with accuracy varying sharply depending on the notation or prompt format, indicating \textit{instruction drift}. Specialized DSP and neural pitch baselines achieve mean scores of
approximately 73.2\% and 75.2\%, respectively, compared with approximately 53.9\% for the strongest ALM.
We release PitchBench as an evaluation benchmark\footnote{
\url{https://huggingface.co/datasets/vaclis/PitchBench}
}, a training dataset, and a Python package\footnote{
\url{https://github.com/vaclisinc/PitchBench}, \url{https://pypi.org/project/pitchbench/}} with data-generation tools to support future research on pitch-aware audio-language modeling.
\end{abstract}

\section{Introduction}




Audio-language models (ALMs) extend the capabilities of large language models to the acoustic domain~\citep{ghosh2026audioflamingonext, gemini31pro, qwenteam2026qwen35omnitechnicalreport, openai2024gpt4oaudio}. In music, frontier models have demonstrated promising performance on genre classification, instrument identification, and captioning~\citep{mmaupro, muse, ghosh2025music}, supporting applications in tutoring, transcription, recommendation, and production. Across these applications, a common perceptual prerequisite emerges: the model must reliably hear pitch.

Pitch, the perceptual correlate of a sound's fundamental 
frequency, underlies melody, harmony, and musical structure, and serves as a prerequisite for downstream music understanding tasks such as melody recognition, comparison, and chord identification. To perform the tasks that ALMs are increasingly expected to support, a model must reliably resolve pitch directly from audio across diverse acoustic conditions. Two modes of pitch perception are commonly distinguished: \textit{absolute pitch}, the 
ability to identify a pitch without an external 
reference~\citep{1993AbsoluteP, theusch2009}, and \textit{relative 
pitch}, the ability to identify the interval between two notes~\citep{dowling1978, krumhansl1990}.

Despite its significance, pitch hearing remains poorly evaluated in
existing benchmarks. Prior work primarily evaluates this capability
through higher-level tasks such as tonality
classification, chord identification, or transposition recognition~\citep{mmau, mmaupro, mmar, muse, cmibench}; they do not systematically evaluate pitch hearing under different acoustic conditions and musical settings. 

As a result, it remains unclear why a model might fail at complex tasks like understanding polyphonic music: 
can it not identify pitches altogether, does it lose pitch
information under temporal variation, or is it not able to discern
simultaneous notes in a polyphonic mixture?
In addition, many music understanding evaluations rely on multiple-choice formats, which can obscure failure to correctly identify pitch: models can exploit distractor design to arrive at correct answers without genuine listening~\citep{muchomusic}.

Across the board, current benchmarks only assess pitch hearing at the surface level, leaving a substantial gap
in our understanding of pitch perception in ALMs. We introduce \textbf{PitchBench}, a systematic evaluation suite 
designed to fill this gap. PitchBench decomposes pitch hearing into 
its constituent perceptual components and evaluates it under 
controlled variation of the acoustic factors. The hierarchical organization gives 
researchers and developers a structured account of what a model can and 
cannot yet hear. We make three 
contributions:

\begin{enumerate}

\item \textbf{A hierarchical evaluation benchmark.} PitchBench comprises 
28 tasks organized into three levels: atomic pitch perception 
(single-tone identification), contextual pitch perception (pitch in 
time, note sequences, chords, and different auditory contexts), and polyphonic pitch perception (pitch 
identification in polyphonic textures).

\item \textbf{A systematic empirical analysis.} We evaluate
5 frontier ALMs alongside specialized DSP and neural pitch baselines.
Dedicated baselines substantially outperform current ALMs, whose pitch perception remains highly unreliable.
Our analysis identifies consistent failure modes invisible in existing higher-level benchmarks, including sensitivity to sound source, note duration, response format, and mild acoustic transformations.

\item \textbf{An open generation pipeline.} All stimuli are generated 
deterministically via \texttt{pitchbench.generation}, a Python package 
released alongside the benchmark. It provides reproducible stimulus 
generation across 19 sound sources and 
all experimental conditions, and is designed to support future dataset 
construction, training, and evaluation for pitch-aware audio-language modeling.

\end{enumerate}

\section{Related Work}
\label{sec:related-work}

\subsection{Audio-Language Models}

The applications of audio-language models (ALMs) have rapidly expanded
from simple classification tasks toward instruction following and open-ended reasoning over audio. Models such as
Pengi~\citep{pengi} framed audio tasks as text generation, learning
audio representations through next-token prediction and supporting
both open-ended and closed-ended tasks. Subsequent systems
including LTU, SALMONN, Qwen-Audio, and Audio
Flamingo~\citep{ltu, salmonn, qwenaudio, audioflamingo} extended such models to speech, environmental
sound, and music. Recent frontier systems, such as Gemini, GPT audio, and Qwen-Omni~\citep{gemini31pro, openai2024gpt4oaudio,
qwenteam2026qwen35omnitechnicalreport}, extend this further with
large-scale multimodal training that includes audio. These general
systems are traditionally evaluated on a wide range of tasks spanning
general audio understanding, multimodal reasoning, and conversational ability. 

\subsection{Benchmarks for Audio and Music Understanding}

\paragraph{General audio benchmarks.}
MMAU, MMAU-Pro, AIR-Bench, and MMAR~\citep{mmau, mmaupro, air-bench, mmar} evaluate audio understanding and reasoning across speech, environmental sound, and music. They assess a broad range of capabilities in ALMs, with music forming one component of their evaluation suites.
SonicBench~\citep{sonicbench} evaluates basic acoustic perception, with pitch assessed through binary high/low classification and comparisons of which of two sounds has the higher pitch.

\paragraph{Music-specific benchmarks.}
MuChoMusic~\citep{muchomusic} evaluates music knowledge and
reasoning across multiple-choice questions covering music theory,
style, genre, and cultural context. 
CMI-Bench~\citep{cmibench} adopts open-ended evaluation across MIR
instruction-following tasks including pitch, key, and beat estimation.
However, benchmarks such as AIR-Bench and CMI-Bench still assess
single-note pitch identification using only NSynth's fixed four-second,
single-note setting~\citep{air-bench, cmibench, nsynth2017}.
This offers an incomplete assessment of pitch perception, lacking
systematic evaluation from low-level pitch identification to higher-level
contextual and polyphonic processing under controlled acoustic variation.

MSU-Bench~\citep{msubench} evaluates score-level music understanding
through open-ended question answering over symbolic musical scores,
covering concepts such as articulation, texture, motif development,
and musical form. Unlike audio-grounded evaluation,
MSU-Bench focuses primarily on reasoning over notated musical structure.
MuSP-Bench~\citep{muspbench} pairs scores with performance audio and MIDI
of classical piano and orchestral works, and finds that frontier models
struggle even more with audio than with scores, indicating that ALMs remain
limited in interpretive and long-horizon musical reasoning.
However, it does not diagnose the lower-level perceptual failures
underlying these difficulties.
MUSE~\citep{muse} tests musical perception through human-validated
listening tasks, including transposition detection, chord identification,
and melodic contour discrimination. However, each task contains only
20 trials, and chord identification is limited to a binary major/minor
distinction. This scale and resolution limit systematic analysis of how
acoustic conditions and response formats affect pitch perception.

\paragraph{Relation to MIR and transcription systems.}

Pitch-related tasks have long been studied in music information
retrieval (MIR) and automatic music transcription (AMT). Classical and
modern systems such as pYIN, CREPE, SPICE, Onsets and Frames, Basic Pitch and
MT3~\citep{6853678, crepe, spice, onsetsframes, bittner2022lightweight, mt3} achieve strong
performance on specialized tasks including pitch estimation, melody
extraction, and symbolic transcription. For general-purpose ALMs, however, pitch perception must support more than transcription: models
are expected to understand what they hear and use that perceptual
information to answer questions, follow instructions, and reason about
higher-level musical content. 

Against this background, PitchBench is positioned as a diagnostic evaluation suite for pitch hearing in audio-language models. Unlike prior music benchmarks, it treats pitch as a controlled experimental axis, systematically varying factors such as absolute versus relative pitch, robustness to perturbations, polyphonic interference, and output representation. Unlike MIR systems, it does not evaluate specialized transcription accuracy, but asks whether general-purpose models encode and utilize pitch information in a way that is stable, accessible, and invariant across prompts and notations. 

\section{PitchBench}
In this section, we present PitchBench as simultaneously an evaluation benchmark for pitch hearing in audio-language models (ALMs), a training dataset, and a Python package for stimulus generation, evaluation, and analysis.

\subsection{Benchmark Task Hierarchy}
PitchBench as a benchmark aims to (1) assess pitch perception of audio-language models under different acoustic conditions and (2) better understand their failure modes. The evaluations are structured around a progressive hierarchy of 28 pitch perception tasks, moving from isolated pitch hearing through time-grounded 
understanding of sounds to polyphonic musical understanding  (Figure~\ref{fig:pyramid}). We classify these tasks according to three levels of pitch perception, which we define as follows:

\begin{figure}
    \centering
    \includegraphics[width=1\linewidth]{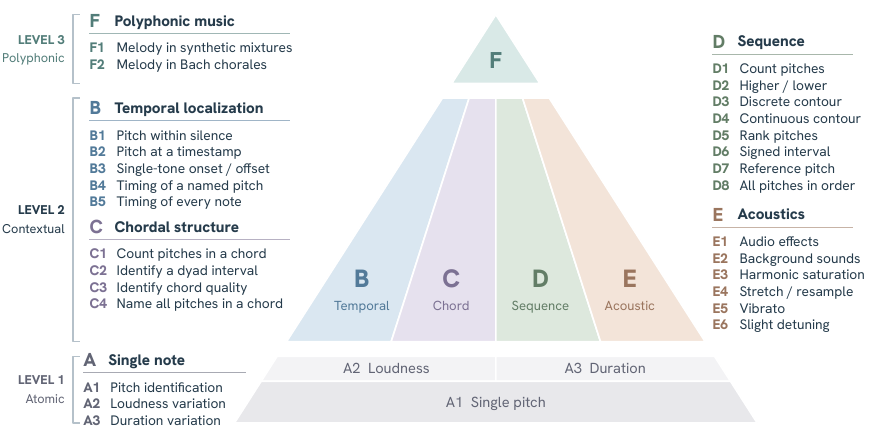}
    \caption{PitchBench: 28 tasks across 3 levels and 6 categories, covering 
    atomic, contextual, and polyphonic pitch perception.}
    \label{fig:pyramid}
\end{figure}

\paragraph{Level 1: Atomic pitch perception.}
At the lowest level, models identify single pitches in audio clips 
containing a single tone for their full duration. These tasks isolate 
core frequency identification under controlled variations in loudness and note duration:
\texttt{A1} identifies a single pitch in a five-second, single-instrument
clip and assesses a more complete range of pitches; \texttt{A2} repeats the task across different loudness levels; and
\texttt{A3} varies note duration from very short to sustained tones, each for a selection of pitches.

\paragraph{Level 2: Contextual pitch perception.}
At the intermediate level, pitch must be identified within the context of time and structure. Category \texttt{B} tests temporal localization; \texttt{C} tests simultaneous pitch relationships; \texttt{D} tests pitch relationships across time; and \texttt{E} tests robustness to acoustic variations. Figure~\ref{fig:pyramid} lists the individual tasks, with detailed descriptions provided in Appendix~\ref{app:contextual-tasks}.

\paragraph{Level 3: Polyphonic pitch perception.}
At the highest level, PitchBench tests whether models can follow a melodic line in polyphonic settings. \texttt{F1} asks models to identify every note belonging to a designated
voice in controlled two- or three-voice synthetic mixtures.
\texttt{F2} extends this setting to real musical structure by asking
models to identify every note belonging to a designated voice in excerpts from
four-part chorales by Johann Sebastian Bach, sourced from the
\texttt{music21} corpus~\citep{cuthbert2010music21}.

Together, these levels probe both absolute and relative pitch 
perception. Categories \texttt{C} and \texttt{D} primarily target relative 
pitch: whether the model can determine which of two notes is higher, 
identify the interval between them, recognize chord quality, and track melodic contour. Categories \texttt{A}, \texttt{B}, and \texttt{E} 
focus on absolute pitch: whether the model can identify a specific pitch 
value regardless of duration, temporal position, or acoustic 
variation.

\subsection{Stimulus Generation}
We use generated stimuli for the following reasons. First, nominal ground truth is known by construction: 
pitch, onset, duration, and all acoustic parameters are specified 
explicitly. Second, synthesis scales 
arbitrarily across the full pitch range, all instruments, and all 
parameter combinations. Third, modern synthesis tools produce stimuli 
that are perceptually realistic while remaining fully controlled. 

Sound sources span 19 timbres: four analytic waveforms (sine, 
sawtooth, square, triangle), which isolate frequency content from 
instrument-specific characteristics, and 15 General MIDI instruments 
rendered via FluidSynth~\citep{fluidr3gm} (piano, electric keyboard, guitar, flute, trumpet,
trombone, clarinet, oboe, violin, cello, organ, bass, synthesizer lead,
synthesizer pad, and voice), covering the principal instrument families. 
For acoustic condition tasks, stimuli are further processed using 
Pedalboard~\citep{sobot2021pedalboard} to apply effects, harmonic saturation, and time-stretching.
Background interference conditions use additive synthetic noise 
textures and real-world recordings.




\subsection{Response Formats and Scoring}
\label{sec:score}
Response modalities include individual pitches, pitch sequences,
counts, and timestamps, all of which can be automatically verified against known ground truth. To avoid privileging models trained on a specific notational convention, each task requiring a pitch label supports three representations: 
\textbf{MIDI}: Note number between 0 and 127  (e.g., \texttt{67}); \textbf{SPN}: Scientific pitch notation, expressed as a 
    letter, an optional accidental, and the octave number 
    (e.g., \texttt{F\#4}); \textbf{Hz}:   Note frequency in Hertz (e.g., \texttt{369.99}).
We additionally collect fixed-do solf\`ege responses (e.g., \texttt{fa\#})
but exclude them from scoring, as they do not encode octave register.


Most tasks are scored by exact accuracy after normalization. For tasks~\texttt{D8}, ~\texttt{F1} and~\texttt{F2}, 
the exact full-sequence match being strict and low-signal, we use an order-aware note-level F1 score.

Let $N_{\mathrm{gt}}$ and $N_{\mathrm{pred}}$ denote the lengths of the
ground-truth and predicted note sequences, respectively, and let $M$
be the length of their longest common subsequence (LCS). We define $P=M/N_{\mathrm{pred}}$,
$R=M/N_{\mathrm{gt}}$, and
$F_1=2PR/(P+R)=2M/(N_{\mathrm{gt}}+N_{\mathrm{pred}})$. Empty predictions receive an F1 score of zero when the
ground-truth sequence is nonempty.
This gives partial credit for correctly recovered notes while remaining sensitive to sequence order: extra predictions reduce precision, missing notes reduce recall, and ordering errors reduce the length of the LCS.

For pitch expressed in Hertz, we apply a 2.5\% relative error margin for voice and 1\% for all other sources. The wider margin for voice accounts for measured tuning offsets of up to approximately 2.4\% between the synthesized voice samples and their nominal target frequencies. For timestamps, we use a 250 millisecond window.
We report complementary distance-based pitch metrics in Appendix~\ref{app:distance-based-metrics}.


\subsection{Specialized Pitch Baselines}

We compare two frozen specialized pitch-estimation baselines on all 28 PitchBench tasks.

\begin{itemize}
    \item \textbf{DSP baseline}: We use YIN~\citep{decheveigne2002yin} for continuous monophonic
fundamental-frequency estimation (30--2500~Hz, 256-ms window,
16-ms hop) and, for simultaneous pitches, harmonic summation over a
constant-Q transform (CQT)~\citep{brown1991cqt}, scoring each candidate
pitch by the $1/h$-weighted magnitudes of its first four harmonics.
The CQT uses 288 bins spanning eight octaves from A0
(36 bins per octave, or three bins per semitone) with a 16-ms hop.
Both are computed with librosa~\citep{mcfee2015librosa}.
    \item \textbf{Neural baseline}: We use Basic Pitch~\citep{bittner2022lightweight} for note-event and
pitch-bend estimation, with a pitch resolution of three bins per
semitone.
\end{itemize}

Because these systems only output pitch and note events, we use rule-based adapters for each task to convert their outputs into the required benchmark formats, such as distinct-pitch counts and intervals.

\subsection{Diagnostic Analyses}
\label{sec:ablations}
We investigate the following research questions:

\begin{itemize}
    \item \textbf{Q1: Is every pitch perceived equally well?}
    For each model, we compute accuracy separately for each ground-truth pitch using the same 3,121 recordings from the 12 single-pitch tasks (\texttt{A1--A3}, \texttt{B1--B2}, \texttt{D7}, and \texttt{E1--E6}).

    \item \textbf{Q2: Does multiple-choice prompting improve pitch
    perception?}
    We use 42 recordings spanning seven pitches (F1, D2, B2, G\#3, F4, D5, B5) and six sound sources (sine wave, synthesizer lead, cello, flute, piano, and voice). We reformulate the pitch-identification questions for these recordings as five-option multiple-choice questions (MCQs). The five options are presented in ascending pitch order, with adjacent options separated by 2, 4, or 6 semitones. We place the correct answer at a randomly selected position among the five options.

\item \textbf{Q3: Is pitch hearing robust to duration, loudness,
temporal placement, and acoustic variation?}
We use the same seven pitches and six sound sources specified in Q2
across all analyses. We evaluate pitch-identification accuracy across
note durations from 50~ms to 60~s, amplitude adjustments from $-30$ to
$+12$~dB, and onset positions at 10, 30, and 50~s within the audio clip.
We also assess the effects of audio processing, background sounds,
harmonic saturation, time stretching, and detuning. For the time-stretching diagnostic, we halve or double each note's
duration using pitch-preserving stretching or resampling, yielding
168 recordings. Resampling shifts the target pitch by one octave.
Table~\ref{tab:pitchbench_unified} gives the number of recordings
per condition.

    \item \textbf{Q4: How does pitch notation affect pitch
    perception?}
    We compare mean scores for MIDI, SPN, and Hz on the same 3,782 recordings from 16 pitch-output tasks: the 12 tasks in Q1 plus \texttt{C4}, \texttt{D8}, \texttt{F1}, and \texttt{F2}. Each recording has equal weight. Scoring follows Table~\ref{tab:pitchbench_transposed}: accuracy for the 12 single-pitch tasks and \texttt{C4}, and order-aware note-level F1 for \texttt{D8}, \texttt{F1}, and \texttt{F2}.
    This analysis assesses instruction following in ALMs and their
    potential biases toward pitch representations used in their
    training data.

    \item \textbf{Q5: Is measured pitch hearing robust to instruction wording?}
    We evaluate all 30 prompts and all five models on a fixed subset of 183 \texttt{A1} recordings, comprising three instrument recordings for each of the 61 pitches. The prompts vary in wording,
    response constraints, and requested notation, ranging from open-ended
    questions to tightly specified answer formats. We report mean accuracy,
    standard deviation, and coefficient of variation across prompts,
    together with 95\% confidence intervals, to quantify variability
    associated with the instruction and response format.
\end{itemize}
\subsection{PitchBench as a Python package}

We release PitchBench as a Python package with standardized run and preview interfaces. Each run records configuration values, prompts, per-item results, plots, and runtime metadata. The generation pipeline, evaluation scripts, and task configurations are provided in our repository.

\section{Results}

We assess five frontier audio-language models (ALMs) that support audio input: Gemini 3.1 Pro~\citep{gemini31pro}, Gemini 3 Flash~\citep{gemini3flash}, 
Qwen-3.5 Omni Plus, Qwen-3.5 Omni Flash~\citep{qwenteam2026qwen35omnitechnicalreport}, 
and Audio Flamingo Next Instruct~\citep{ghosh2026audioflamingonext}.

\begin{table}[!htbp]
\centering
\caption{PitchBench results (\%). \texttt{D8}, \texttt{F1}, and \texttt{F2} use LCS-based order-aware note-level F1; all other tasks use accuracy. Best displayed value per row is bold and second-best underlined (including ties). Mean is the unweighted average of 28 unrounded task scores.}
\label{tab:pitchbench_transposed}
\small
\setlength{\tabcolsep}{3pt}
\resizebox{\textwidth}{!}{%
\begin{tabular}{ll|c|cc|cc|cc}
\toprule
\textbf{Group} & \textbf{Task} & \textbf{Nvidia} & \multicolumn{2}{c|}{\textbf{Google}} & \multicolumn{2}{c|}{\textbf{Alibaba}} & \multicolumn{2}{c}{\textbf{Baselines}} \\
& & \textit{AF-Next-Instruct} & \textit{Gemini 3.1 Pro} & \textit{Gemini 3 Flash} & \textit{Qwen-3.5 Omni Plus} & \textit{Qwen-3.5 Omni Flash} & \textit{DSP} & \textit{Basic Pitch} \\
\midrule
\multirow{3}{*}{A} & A1 \textit{Pitch identification} & 36.0 & 14.9 & 6.0 & 91.6 & 75.1 & \underline{95.8} & \textbf{98.7} \\
 & A2 \textit{Loudness variation} & 43.0 & 22.5 & 23.0 & 90.5 & 71.5 & \underline{97.0} & \textbf{99.0} \\
 & A3 \textit{Duration variation} & 30.5 & 20.0 & 21.0 & 74.8 & 61.0 & \textbf{82.9} & \underline{82.4} \\
\midrule
\multirow{5}{*}{B} & B1 \textit{Pitch within silence} & 5.6 & 25.6 & 21.3 & 90.0 & 76.2 & \underline{98.1} & \textbf{99.4} \\
 & B2 \textit{Pitch at a timestamp} & 10.0 & 17.3 & 16.7 & 77.3 & 56.7 & \underline{94.0} & \textbf{99.3} \\
 & B3 \textit{Single-tone onset / offset} & 0.0 & 0.0 & 2.5 & 13.1 & 9.4 & \underline{78.8} & \textbf{81.3} \\
 & B4 \textit{Timing of a named pitch} & 0.0 & 0.0 & 1.7 & 20.0 & 2.5 & \textbf{65.8} & \underline{51.7} \\
 & B5 \textit{Timing of every note} & 0.0 & 0.0 & 0.0 & \underline{30.0} & 2.5 & \textbf{54.2} & 22.5 \\
\midrule
\multirow{4}{*}{C} & C1 \textit{Count pitches in a chord} & 13.6 & \underline{46.5} & 39.9 & 9.6 & 20.2 & 28.1 & \textbf{47.8} \\
 & C2 \textit{Identify a dyad interval} & 7.8 & 6.9 & 9.1 & 9.1 & 11.2 & \underline{30.2} & \textbf{62.1} \\
 & C3 \textit{Identify chord quality} & 9.9 & 13.0 & 10.4 & 13.0 & 13.0 & \underline{58.9} & \textbf{77.6} \\
 & C4 \textit{Name all pitches in a chord} & 0.0 & 1.5 & 0.0 & 12.0 & 6.5 & \underline{34.0} & \textbf{61.0} \\
\midrule
\multirow{8}{*}{D} & D1 \textit{Count pitches} & 26.4 & 51.4 & 25.0 & \textbf{82.1} & \underline{78.6} & 76.4 & 42.9 \\
 & D2 \textit{Higher / lower} & 51.5 & 71.2 & 50.0 & 72.7 & 65.2 & \textbf{98.5} & \underline{81.8} \\
 & D3 \textit{Discrete contour} & 0.0 & 15.0 & 1.1 & 15.6 & 12.8 & \textbf{61.7} & \underline{56.7} \\
 & D4 \textit{Continuous contour} & 0.0 & 43.8 & 0.0 & \underline{51.9} & 17.5 & \textbf{73.8} & \textbf{73.8} \\
 & D5 \textit{Rank pitches} & 4.2 & 5.0 & 7.5 & 20.0 & 2.5 & \textbf{88.3} & \underline{67.5} \\
 & D6 \textit{Signed interval} & 3.8 & 19.5 & 5.1 & 15.7 & 5.5 & \underline{96.2} & \textbf{97.5} \\
 & D7 \textit{Reference pitch} & 13.8 & 58.5 & 35.4 & 96.2 & 76.9 & \underline{98.5} & \textbf{100.0} \\
 & D8 \textit{All pitches in order} & 14.3 & 18.2 & 14.3 & 86.5 & 48.0 & \underline{94.6} & \textbf{99.6} \\
\midrule
\multirow{6}{*}{E} & E1 \textit{Audio effects} & 35.0 & 21.3 & 19.6 & 86.7 & 55.8 & \textbf{94.6} & \underline{91.3} \\
 & E2 \textit{Background sounds} & 12.1 & 4.2 & 14.2 & \underline{42.9} & 22.9 & 36.7 & \textbf{88.8} \\
 & E3 \textit{Harmonic saturation} & 45.0 & 14.2 & 24.2 & 86.7 & 69.2 & \textbf{97.5} & \underline{96.7} \\
 & E4 \textit{Stretch / resample} & 34.9 & 18.2 & 9.4 & 88.5 & 79.2 & \underline{96.9} & \textbf{99.5} \\
 & E5 \textit{Vibrato} & 24.4 & 2.5 & 3.1 & 65.6 & 39.4 & \textbf{85.6} & \underline{68.8} \\
 & E6 \textit{Slight detuning} & 48.1 & 13.1 & 16.9 & \underline{86.3} & 75.6 & \textbf{98.1} & 73.1 \\
\midrule
\multirow{2}{*}{F} & F1 \textit{Melody in synthetic mixtures} & 26.3 & 16.4 & 30.6 & \underline{35.5} & 30.5 & 20.6 & \textbf{58.8} \\
 & F2 \textit{Melody in Bach chorales} & 26.8 & 14.2 & \underline{31.9} & \textbf{46.4} & 28.7 & 14.7 & 27.3 \\
\midrule
\multicolumn{2}{l|}{\textbf{Mean}} & 18.7 & 19.8 & 15.7 & 53.9 & 39.8 & \underline{73.2} & \textbf{75.2} \\
\bottomrule
\end{tabular}%
}
\end{table}

We report the per-task scoring as described in Section~\ref{sec:score} as well as the mean score across tasks in Table~\ref{tab:pitchbench_transposed}. 
For tasks with pitch-format responses, we select the highest score across MIDI, SPN, and Hz for each item before averaging across items. For accuracy-based tasks, an item is correct if any of the three formats is correct; for D8, F1, and F2, we take the maximum order-aware note-level F1. The overall score is the equally weighted mean of all 28 unrounded task scores.

The results show that specialized baselines substantially outperform all evaluated ALMs, with mean scores of 73.2\% for the DSP pipeline and 75.2\% for
Basic Pitch, suggesting that \textbf{specialized pitch estimation solves many tasks that remain challenging for current ALMs}. Among the ALMs themselves, we also observe substantial differences across model families, both in overall performance and in the pitch information they capture.

\textbf{Qwen-3.5 Omni Plus} achieves the highest mean score among ALMs (53.9\%), performing strongly on sequence counting (82.1\%), reference pitch (96.2\%), stretch (88.5\%), and saturation (86.7\%). However, it struggles with simultaneous pitches in category C. Extracting melodic lines from polyphonic mixtures remains challenging: every ALM scores below 50\% order-aware note-level F1 on both F1 and F2.

\textbf{Qwen-3.5 Omni Flash} is the next strongest ALM (39.8\%). It largely mirrors the Plus model, retaining relatively strong performance on structured tasks while degrading more sharply on complex temporal and compositional reasoning.

\textbf{Gemini 3.1 Pro} (19.8\%) performs unevenly: it is relatively strong in chord counting (C1; 46.5\%), sequence counting (D1; 51.4\%), and reference pitch (58.5\%), but scores near zero on some vibrato and temporal tasks.

\textbf{Audio Flamingo Next Instruct} (18.7\%) performs better on single-pitch identification (\texttt{A1}; 36.0\%) and pitch identification across loudness levels (\texttt{A2}; 43.0\%) than on sequence-based and multi-event tasks, where scores are often near zero.

\textbf{Gemini 3 Flash} (15.7\%) shows low scores on many evaluated tasks.

\begin{figure}
    \centering
    \includegraphics[width=1\linewidth]{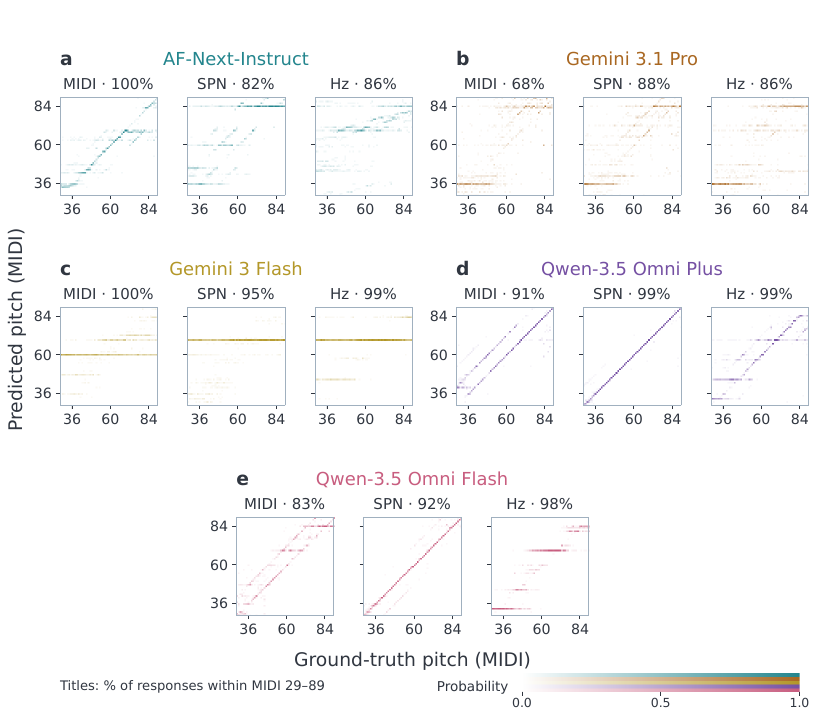}
    \caption{Predicted vs.\ ground-truth note (in MIDI notation) on task \texttt{A1} across three pitch representations.
             Each column is normalized over valid predictions within MIDI~29--89.
             A clean diagonal indicates correct identification; Hz responses are
             rounded to the nearest semitone.
             Percentages above each panel report the fraction of responses within MIDI~29--89.}
    \label{fig:heatmap}
\end{figure}

Figure~\ref{fig:heatmap} shows the per-note prediction distributions for
task \texttt{A1}, which probes each model on single-pitch identification. Model behavior diverges
substantially: some models collapse to a small set of favored pitches
regardless of the input, while others, such as Qwen-3.5 Omni Plus in SPN, track
the ground-truth note closely across the full register, consistent with
its 91.6\% accuracy on this task. Some models tend to confuse the octave of the note, as evidenced by the parallel diagonals exactly 12 semitones, the distance of an octave, apart.

The reference-pitch task (\texttt{D7}) uses a subset of the target pitches and sound sources evaluated in \texttt{A1}. To compare performance on the same notes, we match each D7 trial to the A1 recording with the same target pitch and sound source. Both Gemini models improve in this matched comparison (+44.6 and +28.5 points; Table~\ref{tab:d7-a1-matched}), while both Qwen models decline by 1.5 points and Audio Flamingo Next Instruct declines substantially ($-22.3$), so the benefit of a reference tone is model-dependent.

\section{Diagnostic Analyses}
\label{sec:diagnostic-results}
We answer the five research questions introduced in
Section~\ref{sec:ablations}.

\begin{figure}[h]
    \centering
    \includegraphics[width=1\linewidth]{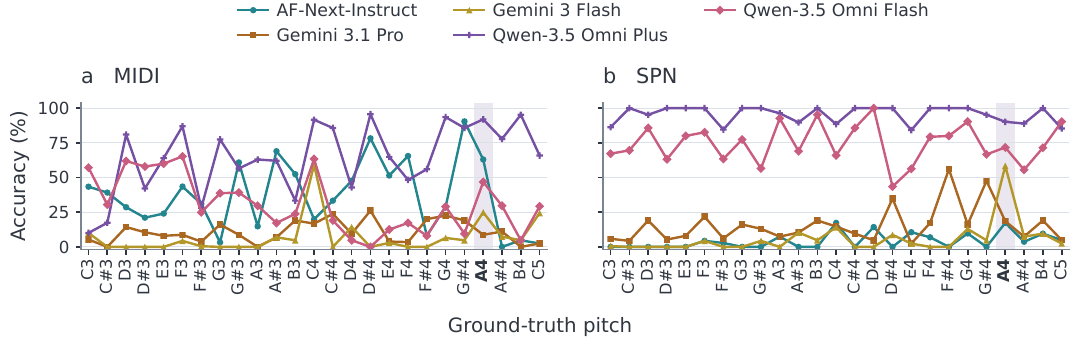}
    \caption{Pitch-wise accuracy across the same 12 single-pitch tasks for every model, shown separately for each ground-truth pitch. For readability, only C3 to C5 are displayed. Left: MIDI notation. Right: SPN notation. Shading marks A4.}
    \label{fig:pitch-comparison}
\end{figure}

\paragraph{Q1: Is every pitch perceived equally well?}
We might expect to see uniformly distributed errors across pitch: equally good or equally bad hearing for every note. As Figure~\ref{fig:pitch-comparison} shows, this does not hold. In MIDI notation, Gemini 3 Flash achieves higher mean pitch-wise accuracy on natural notes (10.4\%) than on sharp or flat notes (2.0\%) within the displayed C3--C5 range. Audio Flamingo Next Instruct shows the opposite pattern in MIDI notation, achieving higher mean pitch-wise accuracy on sharp or flat notes (43.1\%) than on natural notes (32.9\%). This pattern reverses in SPN notation, where its accuracy is higher on natural notes (6.8\%) than on sharp or flat notes (0.7\%). In SPN notation, Qwen models substantially outperform the other models, suggesting a strong model-specific preference for the SPN (letter-plus-octave) response format compared with MIDI format.

A4 (MIDI 69) corresponds to 440~Hz, commonly used as a tuning reference. It is the most frequently predicted note for Gemini 3 Flash in SPN and Hz notation, accounting for 74\% and 78\% of its in-range responses, respectively (Figure~\ref{fig:heatmap}). Among in-range Hz responses rounded to the nearest semitone, A4 is also the most frequently predicted note for Audio Flamingo Next Instruct and both Qwen models. These biases suggest that models' pitch predictions are shaped not only by acoustic evidence from audio, but also by notational priors likely inherited from their language training.

\paragraph{Q2: Does multiple-choice prompting improve pitch perception?}
As shown in Table~\ref{tab:pitchbench_mcq}, multiple-choice questions substantially inflate measured accuracy for models with weak open-ended pitch identification. Gemini 3.1 Pro shows the largest increase, rising from 7.1\% in the
open-ended setting to an average of 58.7\% across MCQ conditions
(+51.6 percentage points). Gemini 3 Flash similarly improves by 35.7 points. In contrast, Qwen-3.5 Omni Plus, already near ceiling at 90.5\%, changes little (+0.8 points), while Qwen-3.5 Omni Flash decreases slightly (-3.2 points).

These gaps suggest that some models may possess partial pitch information that is insufficient for open-ended answering but can still be exploited when candidate labels are provided. Chance may also contribute: uniform random guessing yields 20\% accuracy per five-option question, while the chance level of the reported ANY score depends on the correlation of guesses across MIDI, SPN, and Hz. Option spacing also has a model-dependent effect: accuracy does not consistently increase across models as the spacing between adjacent options increases, suggesting that models do not uniformly exploit pitch proximity to guide selection. Overall, these results motivate us to retain open-ended evaluation as the primary PitchBench setting, as it requires models to predict the pitch label without relying on candidate answers.

\begin{table}[h]
\centering
\caption{PitchBench MCQ: accuracy (\%) for 5-option pitch identification
by spacing between adjacent options in semitones. MCQ mean averages the three spacing
conditions; $\Delta$ is the change from open-ended accuracy in percentage points.}
\label{tab:pitchbench_mcq}
\small
\setlength{\tabcolsep}{4pt}
\resizebox{\textwidth}{!}{%
\begin{tabular}{ll|c|cc|cc}
\toprule
\textbf{Group} & \textbf{Task}
  & \multicolumn{1}{c|}{\textbf{Nvidia}}
  & \multicolumn{2}{c|}{\textbf{Google}}
  & \multicolumn{2}{c}{\textbf{Alibaba}} \\
 &
  & \textit{AF-Next-Instruct}
  & \textit{Gemini 3.1 Pro}
  & \textit{Gemini 3 Flash}
  & \textit{Qwen-3.5 Omni Plus}
  & \textit{Qwen-3.5 Omni Flash} \\
\midrule
 A1 (baseline) & clean signal (open-ended) & 33.3 & 7.1 & 2.4 & 90.5 & 71.4 \\
\midrule\midrule
  \multirow{3}{*}{A1 MCQ} & 2-semitone spacing & 35.7 & 66.7 & 35.7 & 90.5 & 73.8 \\
  & 4-semitone spacing & 35.7 & 57.1 & 38.1 & 95.2 & 69.0 \\
  & 6-semitone spacing & 47.6 & 52.4 & 40.5 & 88.1 & 61.9 \\
\midrule
 & MCQ mean
 & 39.7 & 58.7 & 38.1 & 91.3 & 68.3 \\
 & \(\Delta\) vs. open-ended
 & +6.3 & \textbf{+51.6} & +35.7 & +0.8 & -3.2 \\
\bottomrule
\end{tabular}%
}
\end{table}

\paragraph{Q3: Is pitch hearing robust to duration, loudness, temporal placement, and acoustic variation?}
Table~\ref{tab:pitchbench_unified} reveals a striking duration effect:
Audio Flamingo Next Instruct peaks at 4~s (59.5\%), but its accuracy drops
sharply when duration changes, falling to 19.0\% at 1~s and 28.6\%
at 15~s and 26.2\% at 60~s. At 50 and 250~ms, its accuracy is zero.
As discussed in Section~\ref{sec:related-work}, AIR-Bench and CMI-Bench
assess pitch using NSynth's four-second, single-note
clips~\citep{air-bench, cmibench, nsynth2017}. NSynth is also explicitly
included in Audio Flamingo Next Instruct's training
mixture~\citep{ghosh2026audioflamingonext}. Its strong performance at
4~s and sharp decline outside that duration raise two concerns.
First, evaluation concentrated at a duration represented in training
may overstate general pitch ability. Second, the duration sensitivity
suggests that the learned pitch behavior does not generalize adequately
across temporal variation, motivating more diverse pitch examples during
training. Loudness changes generally have more modest effects, although robustness
varies across models; temporal placement also affects some models.

Detuning effects are model-dependent. Under slight detuning,
Audio Flamingo Next Instruct drops from 33.3\% to 15.5\%, whereas Qwen-3.5 Omni Plus (90.5\% to 85.7\%) and Qwen-3.5 Omni Flash (71.4\% to 73.8\%)
change little. The perturbations leave the nearest standard pitch unchanged.
This matters for real musical audio: a piano may be out of tune.

\begin{table}[h]
\centering
\caption{Unified PitchBench: accuracy (\%) across loudness, temporal structure,
audio effects, and background conditions. $N$ is the number of recordings per row for each model. ANY counts a recording as correct when at least one MIDI, SPN, or Hz response is correct. Means equally weight the 29 condition rows using unrounded accuracies, including the clean-signal baseline. Within each task group, bold
marks the best condition for each model and underline marks the worst.}
\label{tab:pitchbench_unified}
\small
\setlength{\tabcolsep}{4pt}
\resizebox{\textwidth}{!}{%
\begin{tabular}{llr|c|cc|cc}
\toprule
\textbf{Group} & \textbf{Task} & $N$
  & \multicolumn{1}{c|}{\textbf{Nvidia}}
  & \multicolumn{2}{c|}{\textbf{Google}}
  & \multicolumn{2}{c}{\textbf{Alibaba}} \\
 & &
  & \textit{AF-Next-Instruct}
  & \textit{Gemini 3.1 Pro}
  & \textit{Gemini 3 Flash}
  & \textit{Qwen-3.5 Omni Plus}
  & \textit{Qwen-3.5 Omni Flash} \\
\midrule
 A1 (Pitch identification) & clean signal & 42 & 33.3 & 7.1 & 2.4 & 90.5 & 71.4 \\
\midrule\midrule
 \multirow{5}{*}{A2 (Loudness variation)} & \(-30\) dB & 42 & 16.7 & \underline{7.1} & \underline{0.0} & \underline{85.7} & \underline{47.6} \\
  & \(-20\) dB & 42 & 21.4 & \underline{7.1} & \underline{0.0} & \textbf{88.1} & 59.5 \\
  & \(-12\) dB & 42 & \textbf{28.6} & \underline{7.1} & \underline{0.0} & \textbf{88.1} & \textbf{64.3} \\
  & \(+6\) dB & 42 & 21.4 & \underline{7.1} & \textbf{7.1} & \textbf{88.1} & \textbf{64.3} \\
  & \(+12\) dB & 42 & \underline{14.3} & \textbf{9.5} & 2.4 & \underline{85.7} & 61.9 \\
\midrule
 \multirow{6}{*}{A3 (Duration variation)} & 50 ms & 42 & \underline{0.0} & 4.8 & \underline{0.0} & \underline{4.8} & \underline{4.8} \\
  & 250 ms & 42 & \underline{0.0} & \underline{2.4} & \underline{0.0} & 64.3 & 16.7 \\
  & 1 s & 42 & 19.0 & 4.8 & \underline{0.0} & 85.7 & 50.0 \\
  & 4 s & 42 & \textbf{59.5} & 4.8 & 2.4 & \textbf{90.5} & 66.7 \\
  & 15 s & 42 & 28.6 & \textbf{11.9} & \textbf{4.8} & \textbf{90.5} & \textbf{73.8} \\
  & 60 s & 42 & 26.2 & 9.5 & 2.4 & \textbf{90.5} & 66.7 \\
\midrule
 \multirow{3}{*}{B1 (Pitch within silence)} & onset at 10 s & 42 & \textbf{7.1} & \underline{7.1} & 2.4 & \underline{83.3} & \textbf{64.3} \\
  & onset at 30 s & 42 & \underline{2.4} & \underline{7.1} & \textbf{4.8} & \textbf{88.1} & 54.8 \\
  & onset at 50 s & 42 & 4.8 & \textbf{9.5} & \underline{0.0} & \textbf{88.1} & \underline{50.0} \\
\midrule
 \multirow{5}{*}{E1 (Audio effects)} & High-pass filtering & 42 & \textbf{23.8} & \underline{4.8} & \underline{0.0} & \underline{76.2} & 42.9 \\
  & Low-pass filtering & 42 & 14.3 & 7.1 & \underline{0.0} & 88.1 & 38.1 \\
  & Distortion & 42 & 14.3 & \underline{4.8} & \underline{0.0} & \textbf{90.5} & 45.2 \\
  & Reverb & 42 & 21.4 & \textbf{16.7} & \textbf{4.8} & 88.1 & \textbf{59.5} \\
  & Chorus & 42 & \underline{9.5} & 11.9 & 2.4 & 78.6 & \underline{33.3} \\
\midrule
 \multirow{6}{*}{E2 (Background sounds)} & white noise & 42 & \underline{0.0} & 2.4 & 2.4 & 26.2 & \textbf{7.1} \\
  & bells & 42 & 4.8 & 7.1 & \underline{0.0} & \underline{2.4} & 2.4 \\
  & crowd noise & 42 & 4.8 & 4.8 & \underline{0.0} & 19.0 & \textbf{7.1} \\
  & rain & 42 & \underline{0.0} & 4.8 & 2.4 & \textbf{35.7} & \underline{0.0} \\
  & street ambience & 42 & 2.4 & \textbf{14.3} & \textbf{4.8} & 14.3 & \underline{0.0} \\
  & oscillating noise & 42 & \textbf{9.5} & \underline{0.0} & 2.4 & 4.8 & \underline{0.0} \\
\midrule
 E3 (Harmonic saturation) & saturation & 126 & 18.3 & 5.6 & 0.8 & 90.5 & 69.0 \\
\midrule
 E4 (Stretch / resample) & time stretching & 168 & 17.3 & 11.3 & 4.2 & 83.3 & 54.8 \\
\midrule
 E6 (Slight detuning) & detuning & 252 & 15.5 & 9.9 & 5.2 & 85.7 & 73.8 \\
\midrule
\multicolumn{3}{l|}{\textbf{Mean}} & 15.1 & 7.3 & 2.0 & 68.8 & 43.1 \\
\bottomrule
\end{tabular}%
}
\end{table}

\paragraph{Q4: How does pitch notation affect pitch perception?}
Table~\ref{tab:notation-comparison} shows model-specific preferences
for output notation. SPN gives the highest mean score among the three
requested formats for Qwen-3.5 Omni Plus (74.0\%), Qwen-3.5 Omni Flash (53.3\%),
and Gemini 3.1 Pro (9.1\%), whereas MIDI performs best for Audio Flamingo Next Instruct (23.5\%) and Gemini 3 Flash (8.2\%). Hz is weakest for Audio Flamingo Next Instruct, Gemini 3 Flash, and
both Qwen models.
These differences show that measured performance depends on the requested
notation. 


\begin{table}[!htbp]
    \centering
    \caption{Mean score (\%) by pitch notation on the same 3,782 recordings
    from 16 pitch-output tasks for every model, using
    Table~\ref{tab:pitchbench_transposed}'s scoring.
    Bold marks each model's best score among the three requested formats.
    ANY takes the highest score across MIDI, SPN, and Hz for each recording.}
    \label{tab:notation-comparison}
    \small
    \setlength{\tabcolsep}{8pt}
    \renewcommand{\arraystretch}{1.15}
    \begin{tabularx}{\linewidth}{
        @{}l
        !{\vrule width 0.8pt}
        *{3}{>{\centering\arraybackslash}X}
        !{\vrule width 0.3pt}
        >{\centering\arraybackslash}X@{}
    }
        \toprule
        \textbf{Model} & \textbf{MIDI} & \textbf{SPN} & \textbf{Hz} & \textbf{ANY} \\
        \midrule
        AF-Next-Instruct     & \textbf{23.5} & 7.8 & 2.2 & 28.0 \\
        Gemini 3.1 Pro       & 6.9 & \textbf{9.1} & 7.1 & 16.4 \\
        Gemini 3 Flash       & \textbf{8.2} & 7.6 & 6.7 & 14.3 \\
        Qwen-3.5 Omni Plus   & 45.0 & \textbf{74.0} & 33.8 & 76.8 \\
        Qwen-3.5 Omni Flash  & 23.1 & \textbf{53.3} & 13.2 & 59.1 \\
        \bottomrule
    \end{tabularx}
\end{table}

\paragraph{Q5: Is measured pitch hearing robust to instruction wording?}
Table~\ref{tab:prompt-sensitivity} in Appendix~\ref{app:prompt-sensitivity}
shows a substantial but
model-dependent effect of prompt form. Across the 30 prompts asking the models to identify a single pitch,
the standard deviation of accuracy ranges from 2.4 to 30.8 percentage
points, with coefficients of variation between 63.1\% and 85.8\%.
Because the same audio recordings are used throughout, this variation
reflects sensitivity to instruction wording and the requested response
form. We refer to this variation in measured performance as
\textit{instruction drift}. It shows that pitch identification is not
stable across the tested prompts and notations. 

\section{Limitations and Future Work}

PitchBench currently relies on algorithmically synthesized target notes, with real-world recordings used for some background interference conditions. While this offers a sound ground truth and ensures precise control over acoustic parameters, the generated stimuli approximate real recordings by design and do not fully capture their timbral complexity and acoustic naturalism. The instruments available in PitchBench (General MIDI instruments via FluidSynth) represent a narrow slice of global musical practice, and we do not extensively probe non-Western instruments, vocal techniques, or timbrally unconventional sound sources that may be common in other musical traditions. Future work extends PitchBench to incorporate real musical recordings and more demanding polyphonic pieces that build on \texttt{F1} and \texttt{F2}, and to broaden the instrument pool and include non-Western tuning systems and repertoire, increasing the cultural and acoustic diversity of the benchmark.

\section{Conclusion}
We introduced PitchBench, a systematic evaluation suite comprising 28 tasks that decompose pitch perception into atomic, contextual, and polyphonic levels. Our evaluation of five frontier ALMs reveals that pitch hearing remains substantially unreliable, with performance varying across acoustic conditions, note durations, and pitch representations. Polyphonic pitch perception remains challenging. By decomposing pitch perception into controlled sub-tasks, PitchBench provides a diagnostic account of \emph{where} models fall short of reliable musical understanding. We hope PitchBench offers a useful foundation for future work on building audio-language models with perfect pitch.

\begin{ack}
Thoughtful Lab, with which two of the authors are affiliated, provided OpenRouter and Claude API credits used for model evaluation in this work. The authors declare no competing interests.
\end{ack}

\bibliographystyle{plainnat}

\bibliography{references}







\newpage
\appendix

\section{Contextual Pitch Perception Tasks}
\label{app:contextual-tasks}
\begin{itemize}
    \item \textbf{\texttt{B} --- Temporal localization.}
    These tasks test whether models can jointly identify pitch and
    its location in time. \texttt{B1} targets identification of a single pitch surrounded by
    silence; \texttt{B2} identifies the pitch sounding at a specified
    timestamp; \texttt{B3} reports the onset and offset of a single tone;
    \texttt{B4} localizes a named pitch in time among distractor tones; and
    \texttt{B5} reports the timing of multiple notes in a sequence.

    \item \textbf{\texttt{C} --- Chordal structure.}
    These tasks test simultaneous pitch relationships.
    \texttt{C1} counts the number of distinct pitches in a chord;
    \texttt{C2} identifies the interval between two simultaneous pitches;
    \texttt{C3} classifies chord quality; and \texttt{C4} enumerates all
    pitches sounding in a chord.

    \item \textbf{\texttt{D} --- Sequential structure.}
    These tasks test pitch relationships across time.
    \texttt{D1} counts pitches in a sequence;
    \texttt{D2} determines which of two successive tones is higher;
    \texttt{D3} describes a discrete pitch trajectory;
    \texttt{D4} describes a continuous pitch trajectory;
    \texttt{D5} ranks notes from lowest to highest;
    \texttt{D6} identifies the signed interval between successive notes;
    \texttt{D7} identifies a target pitch relative to a named reference
    tone, with both tones presented in one audio clip; and
    \texttt{D8} transcribes every pitch in a sequence in order.

    \item \textbf{\texttt{E} --- Acoustic variations.}
    These tasks test robustness to signal-level transformations
    while preserving or systematically perturbing pitch content.
    \texttt{E1} applies audio effects;
    \texttt{E2} introduces background interference;
    \texttt{E3} increases harmonic saturation;
    \texttt{E4} time-stretches or resamples the tone;
    \texttt{E5} introduces pitch instability through vibrato; and
    \texttt{E6} slightly detunes the tone.
\end{itemize}

\section{Distance-Based (Semitone) Metrics}
\label{app:distance-based-metrics}

From a musician's perspective, an incorrect pitch can alter the resulting
harmony or lead to an incorrect chord interpretation. This can occur whether
the error is one semitone or one octave. We therefore use exact-pitch
accuracy as the primary evaluation metric.

To provide a more fine-grained performance signal, we additionally report
distance-aware diagnostics for the 12 tasks whose answer is one absolute
pitch (\texttt{A1--A3}, \texttt{B1--B2}, \texttt{D7}, and
\texttt{E1--E6}). For target pitch $y$ and prediction $\hat{y}$ on the
continuous MIDI scale, the absolute error in semitones is
\begin{equation}
    d = |\hat{y} - y|.
    \label{eq:semitone-error}
\end{equation}
Because we support multiple notations, we calculate these diagnostics using the prediction closest to the ground truth across MIDI, SPN, and Hz, analogous to \emph{Any} accuracy. Hz predictions are converted to continuous MIDI values.

\textbf{Mean absolute error (MAE)} is the mean of $d$, and
\textbf{median error} is its median. The \textbf{$\leq 1$-semitone rate}
is the fraction of predictions with $d \leq 1$.
\textbf{Pitch-class accuracy} ignores octave, whereas \textbf{octave-only error} is the fraction of predictions with the correct pitch class but the wrong octave. These two metrics use the selected prediction rounded to the nearest semitone.

All summaries give the 12 tasks equal weight. MAE is the mean of the task-level MAEs; the median is a weighted median in which each task contributes equal total weight, divided equally among its valid predictions. Percentage metrics use all recordings as the denominator; unparseable responses satisfy none of the correctness or octave-only-error conditions. MAE and median error exclude unparseable responses. Each model has at least one valid MIDI, SPN, or Hz prediction for all 3,121 recordings, except Gemini 3.1 Pro, which has valid predictions for 3,120.

\newsavebox{\disttabbox}
\newlength{\appendixtablewidth}
\begin{table}[!htbp]
\centering
\caption{Exact accuracy and distance-aware diagnostics for the 12
single-absolute-pitch tasks (\texttt{A1--A3}, \texttt{B1--B2},
\texttt{D7}, and \texttt{E1--E6}). All summaries give the 12 tasks equal
weight. Exact accuracy counts a recording as correct if any MIDI, SPN,
or Hz response is correct. Distance metrics use the closest prediction
across these formats. MAE and the weighted median use valid predictions;
percentage metrics use all recordings as the denominator. Arrows indicate
the preferred direction; bold denotes the best displayed value for each
metric, including ties.}
\label{tab:distance-based-metrics}
\footnotesize
\setlength{\tabcolsep}{3pt}
\begin{lrbox}{\disttabbox}
\begin{tabular}{lrrrrr}
\toprule
\textbf{Metric} &
\shortstack{\textbf{AF-Next-}\\\textbf{Instruct}} &
\shortstack{\textbf{Gemini}\\\textbf{3.1 Pro}} &
\shortstack{\textbf{Gemini}\\\textbf{3 Flash}} &
\shortstack{\textbf{Qwen-3.5}\\\textbf{Omni Plus}} &
\shortstack{\textbf{Qwen-3.5}\\\textbf{Omni Flash}} \\
\midrule
Exact accuracy (\%) $\uparrow$ & 28.2 & 19.4 & 17.5 & \textbf{81.4} & 63.3 \\
MAE (semitones) $\downarrow$ & 6.06 & 8.42 & 8.69 & \textbf{1.27} & 2.16 \\
Median error (semitones) $\downarrow$ & 4.00 & 6.00 & 5.00 & \textbf{0.00} & \textbf{0.00} \\
$\leq 1$-semitone rate (\%) $\uparrow$ & 36.3 & 23.5 & 22.3 & \textbf{86.1} & 71.8 \\
Pitch-class accuracy (\%) $\uparrow$ & 31.5 & 32.0 & 26.1 & \textbf{83.1} & 66.4 \\
Octave-only error (\%) $\downarrow$ & 2.6 & 12.6 & 8.5 & \textbf{1.5} & 2.9 \\
\bottomrule
\end{tabular}
\end{lrbox}
\global\appendixtablewidth=\wd\disttabbox
\usebox{\disttabbox}
\end{table}

Table~\ref{tab:distance-based-metrics} provides complementary information
to exact accuracy. Audio Flamingo Next Instruct's exact accuracy is 8.8 percentage
points higher than Gemini 3.1 Pro's (using unrounded scores), while its MAE is 2.36 semitones lower.
Its within-one-semitone rate is also 12.8 percentage points higher,
indicating that its predictions tend to be closer to the correct pitch.
Across the five models in the table, higher exact accuracy is consistently
associated with lower MAE.

\newpage

\section{Effect of a Reference Tone}
\label{app:reference-tone}
The reference-pitch task (\texttt{D7}) presents a named reference tone
followed by a target tone. Every \texttt{D7} target shares its pitch and
sound source with exactly one \texttt{A1} recording, so we compare each
model on the same 130 target notes with and without the reference.
Table~\ref{tab:d7-a1-matched} reports the matched accuracies, counting a
recording as correct if any MIDI, SPN, or Hz response is correct.

\begin{table}[!ht]
\centering
\caption{Accuracy (\%) on the 130 \texttt{D7} trials and the matched
\texttt{A1} recordings; $\Delta$ is \texttt{D7} minus \texttt{A1}.}
\label{tab:d7-a1-matched}
\footnotesize
\setlength{\tabcolsep}{3pt}
\begin{tabular*}{\appendixtablewidth}{@{\extracolsep{\fill}}lrrrrr@{}}
\toprule
 &
\shortstack{\textbf{AF-Next-}\\\textbf{Instruct}} &
\shortstack{\textbf{Gemini}\\\textbf{3.1 Pro}} &
\shortstack{\textbf{Gemini}\\\textbf{3 Flash}} &
\shortstack{\textbf{Qwen-3.5}\\\textbf{Omni Plus}} &
\shortstack{\textbf{Qwen-3.5}\\\textbf{Omni Flash}} \\
\midrule
Matched \texttt{A1} accuracy (\%) & 36.2 & 13.8 & 6.9 & 97.7 & 78.5 \\
\texttt{D7} accuracy (\%) & 13.8 & 58.5 & 35.4 & 96.2 & 76.9 \\
$\Delta$ (points) & $-22.3$ & $+44.6$ & $+28.5$ & $-1.5$ & $-1.5$ \\
\bottomrule
\end{tabular*}
\end{table}

\section{Prompt Sensitivity}
\label{app:prompt-sensitivity}
Table~\ref{tab:prompt-sensitivity} reports the model-wise summary
statistics for the prompt-sensitivity analysis in
Section~\ref{sec:diagnostic-results}. We evaluate all 30 prompts and all five models on a fixed subset of 183 \texttt{A1} recordings, comprising three instrument recordings for each of the 61 pitches.
We compute 95\% percentile-bootstrap confidence intervals using 10,000 resamples of the 183 recordings (seed 42), preserving pairing across prompts and models. The 30 prompts remain fixed. Intervals use the 2.5th and 97.5th percentiles and quantify recording-sampling uncertainty for these prompts. SD is the population standard deviation across the 30 prompt accuracies, in percentage points (pp).

\begin{table}[!htbp]
\centering
\caption{Prompt sensitivity on Task~\texttt{A1}. Each estimate is shown above
its 95\% bootstrap confidence interval; CV is SD divided by mean accuracy.}
\label{tab:prompt-sensitivity}
\small
\setlength{\tabcolsep}{0pt}
\renewcommand{\arraystretch}{1.15}
\newcommand{\promptstat}[3]{\shortstack{#1\\[-1pt]{\scriptsize\textcolor{black!65}{[#2, #3]}}}}
\begin{tabular*}{\appendixtablewidth}{@{\extracolsep{\fill}}lccc@{}}
\toprule
\textbf{Model} & \textbf{Mean accuracy (\%)} & \textbf{SD (pp)} & \textbf{CV (\%)} \\
\midrule
AF-Next-Instruct & \promptstat{24.9}{21.7}{28.0} & \promptstat{19.2}{16.8}{21.9} & \promptstat{77.4}{73.8}{82.0} \\
\addlinespace[5pt]
Gemini 3.1 Pro & \promptstat{6.5}{5.0}{8.2} & \promptstat{5.6}{4.7}{6.9} & \promptstat{85.8}{77.3}{104.7} \\
\addlinespace[5pt]
Gemini 3 Flash & \promptstat{3.8}{2.6}{5.1} & \promptstat{2.4}{1.9}{3.6} & \promptstat{63.1}{52.4}{96.3} \\
\addlinespace[5pt]
Qwen-3.5 Omni Plus & \promptstat{41.1}{38.4}{43.8} & \promptstat{30.8}{29.5}{32.3} & \promptstat{75.1}{69.6}{81.3} \\
\addlinespace[5pt]
Qwen-3.5 Omni Flash & \promptstat{34.0}{31.1}{37.0} & \promptstat{22.5}{21.2}{24.0} & \promptstat{66.1}{60.1}{73.7} \\
\bottomrule
\end{tabular*}
\end{table}

\newpage

\subsection{Prompts used in the evaluation}
\label{app:prompt-texts}
The 30 evaluated prompts are reproduced below in their original order, grouped by wording and response constraints.

\begingroup
\small
\newenvironment{promptlist}[1]{%
  \begin{list}{\textbf{P\ifnum\value{enumi}<10 0\fi\arabic{enumi}}}{%
    \usecounter{enumi}%
    \setlength{\leftmargin}{3em}%
    \setlength{\labelwidth}{2.4em}%
    \setlength{\labelsep}{0.6em}%
    \setlength{\itemsep}{5pt}%
    \setlength{\parsep}{0pt}%
    \setlength{\topsep}{4pt}%
  }%
  \setcounter{enumi}{#1}%
  \raggedright
}{\end{list}}
\newtcolorbox{promptbox}[1]{breakable, enhanced, colback=gray!4, colframe=gray!30,
  boxrule=0.6pt, arc=2mm, left=8pt, right=8pt, top=4mm, bottom=6pt,
  title={#1}, fonttitle=\bfseries\small, coltitle=black,
  attach boxed title to top left={xshift=3mm, yshift=-2mm},
  boxed title style={colback=gray!15, colframe=gray!30, boxrule=0.6pt, arc=1.5mm},
  before skip=4mm, after skip=3mm}
\newcommand{\pchip}[1]{\tcbox[on line, colback=gray!8, colframe=gray!45,
  boxrule=0.4pt, arc=1pt, boxsep=0pt, left=2pt, right=2pt, top=1pt, bottom=1pt]{\strut #1}}

\begin{promptbox}{Sentence completions}
\begin{promptlist}{0}
    \item This pitch is...
    \item This note is...
    \item The MIDI value of this pitch is...
    \item The frequency of this pitch in Hz is...
    \item The solfège name of this pitch is...
    \item The Scientific Pitch Notation of this pitch is...
\end{promptlist}
\end{promptbox}

\begin{promptbox}{Short questions with answer constraints}
\begin{promptlist}{6}
    \item Which pitch is this? Answer only with the pitch, no other text.
    \item Which note is this? Answer only with the note, no other text.
    \item What is the MIDI value for this single note? Answer only with the MIDI value, no other text.
    \item What is the frequency in Hz for this note? Answer only with the frequency in Hz, no other text.
    \item What is the solfège name for this note? Answer only with the solfège name, no other text.
    \item What is the note in Scientific Pitch Notation for this note? Answer only with the note in Scientific Pitch Notation, no other text.
\end{promptlist}
\end{promptbox}

\begin{promptbox}{Detailed task instructions}
\begin{promptlist}{12}
    \item This audio contains a single musical note. What is the MIDI note number (an integer from 0 to 127)? Reply with ONLY the integer. Nothing else. Output only the answer.
    \item This audio contains a single musical note. What is the main pitch frequency, expressed in Hertz? Reply with ONLY a number (the frequency in Hz). Nothing else. Output only the answer.
    \item This audio contains a single musical note. What is the solfege syllable and accidental (if needed) of this pitch? Use fixed-do (do=C, re=D, mi=E, fa=F, sol=G, la=A, si=B). Include sharps or flats when needed (e.g. \pchip{do\#}, \pchip{reb}, \pchip{fa\#}, \pchip{sib}). Reply with ONLY the solfege syllable and accidental (if needed). Nothing else. Output only the answer.
    \item This audio contains a single musical note. What is the note name and octave? Reply with ONLY the note name in Scientific Pitch Notation (e.g. \pchip{C4}, \pchip{F\#3}, \pchip{Bb5}). Nothing else. Output only the answer.
\end{promptlist}
\end{promptbox}

\begin{promptbox}{Alternative wordings}
\begin{promptlist}{16}
    \item What is the pitch of this note?
    \item What pitch is being played?
    \item What note do you hear in this recording?
    \item Listen to the audio and identify the musical note.
    \item Name the fundamental pitch of the sound.
    \item Can you tell which note the instrument is playing?
    \item Transcribe the sustained sound as a note and octave.
    \item What musical tone is sounding in this clip?
    \item If this sound were matched to a piano key, which key would it be?
    \item Identify the pitch. For example, \pchip{C4} or \pchip{F\#5}.
    \item Report the pitch in a standard form, such as \pchip{A4}, \pchip{MIDI 69}, or \pchip{440 Hz}.
    \item Determine the audio’s fundamental and provide its musical pitch.
    \item Estimate the fundamental frequency.
    \item Pitch-label the audio.
\end{promptlist}
\end{promptbox}
\endgroup

\end{document}